\documentclass[preprint,aps,showpacs]{revtex4}
\usepackage{bm}

\begin{document}

\count255=\time\divide\count255 by 60 \xdef\hourmin{\number\count255}
  \multiply\count255 by-60\advance\count255 by\time
 \xdef\hourmin{\hourmin:\ifnum\count255<10 0\fi\the\count255}

\newcommand{\xbf}[1]{\mbox{\boldmath $ #1 $}}

\newcommand{\sixj}[6]{\mbox{$\left\{ \begin{array}{ccc} {#1} & {#2} &
{#3} \\ {#4} & {#5} & {#6} \end{array} \right\}$}}

\newcommand{\threej}[6]{\mbox{$\left( \begin{array}{ccc} {#1} & {#2} &
{#3} \\ {#4} & {#5} & {#6} \end{array} \right)$}}

\title{Hyperon Radiative Decays in the $1/N_c$ Expansion}

\author{Richard F. Lebed}
\email{Richard.Lebed@asu.edu}

\author{Daniel R. Martin}
\email{daniel.martin@asu.edu}


\affiliation{Department of Physics and Astronomy, Arizona State University, 
Tempe, AZ 85287-1504}

\date{April, 2004}


\begin{abstract}
Using a recent calculation of transition magnetic moments in the
$1/N_c$ expansion and a calculation showing the suppression of $E2/M1$
by powers of $N_c$, we compute the widths for the radiative decays
$\Sigma^* \to \Sigma \, \gamma$, $\Sigma^* \to \Lambda \, \gamma$, and
$\Xi^* \to \Xi \, \gamma$.
\end{abstract}


\maketitle
\thispagestyle{empty}

\newpage
\setcounter{page}{1}

In this Report we combine the results of two calculations using the
$1/N_c$ expansion of QCD to predict radiative widths of strange
spin-3/2 baryon resonances.  The first, by the current
authors~\cite{LM}, predicts all unmeasured diagonal and transition
magnetic moments to relative accuracy $\varepsilon^2/N_c^2$, where
$\varepsilon$ is the dimensionless scale of SU(3) flavor breaking
(shown in Ref.~\cite{LM} to be at most $\simeq 1/3$).  The second, by
Jenkins, Ji, and Manohar~\cite{JJM}, shows that $M$ (spin-3/2) $\to m$
(spin-1/2) $\gamma$ decays have an $E2/M1$ amplitude ratio of
$O(1/N_c^2)$.  Only $\Delta \to N \gamma$ is considered in
Ref.~\cite{JJM}, but the same conclusion holds for almost the full
SU(3) baryon multiplets, with the exceptions noted below.  This
conclusion follows because $\Delta J \! = \!  2$ quadrupole operators
require one more quark interaction, and hence an additional factor of
$1/N_c$, than $\Delta J \! = \! 1$ magnetic moment
operators~\footnote{It should be noted that this suppression can in
principle be wiped out if the additional quark contribution adds
coherently for all $N_c$ quarks in the baryon.  For the quadrupole
case, the only operator for which this holds is $G^{iQ} G^{j3}/N_c$,
where $G$ is a combined spin-flavor operator (Eq.~(2.1) of
Ref.~\cite{LM}), $i$ and $j$ are spin indices, $Q$ is the flavor index
corresponding to coupling to a photon according to the quark charges
$(q_u,q_d,q_s)$, and the flavor 3 index indicates a contribution along
the $\Delta I \! = \! 1$, $\Delta I^3 \! = \! 0$ direction.  In fact,
such an operator appears with the usual $O(10^{-3})$ coefficient
characteristic of isospin breaking, and therefore is negligible here.}.

The full radiative width is given by
\begin{equation} \label{width}
\Gamma (M \to m \, \gamma) = \frac{k_\gamma^2}{\pi} \frac{m}{2M}
\left[ |M1|^2 + 3|E2|^2 \right] , 
\end{equation}
where
\begin{equation} \label{M1}
M1 = \frac{e}{2m} k_\gamma^{\, 1/2} \mu^{\vphantom\dagger}_{Mm}
\end{equation}
defines the transition moment $\mu^{\vphantom\dagger}_{Mm}$, and the
photon momentum is
\begin{equation} \label{kin}
k_\gamma = \frac{M^2-m^2}{2M} \ .
\end{equation}

From Eq.~(\ref{width}), one sees that completely neglecting the $E2$
amplitude introduces only a relative $O(1/N_c^4)$ correction to the
width for $O(N_c^1)$ transition magnetic moments.  Since this is
comparable to the effect of the calculated uncertainty of the
transition moments in $|M1|^2$, we include it in the stated
uncertainties; however, the relative $E2$ contribution to the width
($3/N_c^4 \simeq 4 \%$) is somewhat special in that it is positive
definite.  We accommodate this effect by shifting the predicted width
values upward by this amount as well as adding in quadrature an
additional uncertainty of this magnitude: In other words, the central
value chosen for the $E2$ contribution is given by its natural size
according to the $1/N_c$ expansion, and the uncertainty allows it to
range between zero and twice this size.

An exception to this $N_c$ scaling occurs for $\Sigma^{*-} \! \to \!
\Sigma^- \, \gamma$ or $\Xi^{*-} \! \to \! \Xi^- \, \gamma$, which
vanish in the limit of SU(3) symmetry for $N_c \! \to 3$.  Since the
$U$-spin subgroup of SU(3) does not change electric charge, the photon
is a $U$-spin scalar.  On the other hand, for $N_c \! = 3$,
$\Sigma^{*-}$ and $\Xi^{*-}$ belong to a $U \! = \frac 3 2$ quartet,
while $\Sigma^-$ and $\Xi^-$ form a $U \! = \frac 1 2$ doublet, which
forbids the transition in the SU(3) limit.  Although this multiplet
structure is somewhat modified for $N_c \! > \! 3$, the final step of
a calculation in the $1/N_c$ expansion sets $N_c \! = 3$.  Therefore,
the smallness of the corresponding transition moments, due to $U$-spin
conservation, is still respected by the $1/N_c$ expansion: All
SU(3)-conserving contributions to these processes are proportional to
$(N_c \! - 3)$.  In fact, the leading operators for both the
transition magnetic and electric quadrupole moments in these cases
have $O(\varepsilon N_c^0)$ matrix elements~\footnote{The leading
magnetic moment operators are~\cite{LM} $\varepsilon q_s J_s^3$ and
$\varepsilon Q J_s^3/N_c$, where $q_s = -\frac 1 3$ and $J_s$ is the
angular momentum operator acting only on the strange quarks.  The
leading quadrupole operator in this notation is $\varepsilon \{
G^{iQ}, J_s^j \} /N_c$.}, meaning that the only nontrivial scaling of
$E2/M1$ comes from kinematic factors of the photon
momentum~\cite{JJM}, $k_\gamma^{\, 3/2}/k_\gamma^{\, 1/2} \! =
O(1/N_c)$, so that we estimate a relative $3/N_c^2 \! \simeq 33\%$
uncertainty from neglecting the $E2$ width contribution to
$\Sigma^{*-} \! \to \! \Sigma^- \, \gamma$ or $\Xi^{*-} \! \to \!
\Xi^-
\, \gamma$.

Note that we only estimate the quadrupole moments according to their
magnitude and $1/N_c$ scaling; it would also be possible to carry out
a more detailed analysis using, for example, the measured $E2$ \,
$\Delta^+ \! \to p \, \gamma$ transition and the results of a $1/N_c$
analysis among the quadrupole moments~\cite{quad} to determine the
others.  However, since the $E2$ contributions are seen to be smaller
than $M1$ in all cases---often much smaller---this refinement is not
essential for our purposes.

Combining Eqs.~(\ref{width}), (\ref{M1}), and (\ref{kin}), and using
the predictions in Table~XI of Ref.~\cite{LM} gives the results in
Table~\ref{pred}.  While a number of calculations to predict these
widths have been carried out in the past (using such varied methods as
quark models~\cite{quark}, the MIT bag model~\cite{bag}, Skyrme
models~\cite{Skyrme}, algebraic models~\cite{alg}, heavy baryon chiral
perturbation theory~\cite{HBT}, and lattice QCD~\cite{lat}), we point
out that this calculation is not only model independent, but depends
upon experimental input only from the 11 measured magnetic
moments~\cite{LM}, and upon expansions in $1/N_c \!  = 1/3$ and
$\varepsilon \alt 1/3$.

\begin{table}
\caption{Predicted values of the radiative widths (in keV).
\label{pred}}
\begin{tabular}{ll|ll|ll}
\hline\hline
$\Sigma^{*+} \! \! \to \Sigma^+ \, \gamma$ & \ $118 \pm 10$ \ & \
$\Sigma^{*0} \! \! \to \Lambda  \, \gamma$ & \ $298 \pm 25$ \ & \
$\Sigma^{*0} \! \! \to \Sigma^0 \, \gamma$ & \ $24.9 \pm 4.1$ \\
$\Sigma^{*-} \! \! \to \Sigma^- \, \gamma$ & \ $0.58 \pm 0.70$ \ & \
$\Xi^{*0}    \! \! \to \Xi^0    \, \gamma$ & \ $135 \pm 12$ \ & \
$\Xi^{*-}    \! \! \to \Xi^-    \, \gamma$ & \ $0.68 \pm 0.82$ \\
\hline
\end{tabular}
\end{table}

\vskip 1em
{\em Acknowledgments}.
We thank Simon Taylor for discussions regarding previous width
calculations.  This work was supported in part by the National Science
Foundation under Grant No.\ PHY-0140362.

\end{document}